\documentclass[aps,prl,showpacs,twocolumn,preprintnumbers,amsmath,amssymb]{revtex4}
\usepackage{graphicx} 
\usepackage{dcolumn} 
\usepackage{bm} 

\begin{document}
\title{Glassy dynamics in thermally-activated list sorting}
\author{Ling-Nan Zou}\email{zou@uchicago.edu}
\author{Sidney R. Nagel}\email{srnagel@uchicago.edu}
\affiliation{The James Franck Institute and Department of Physics, The University of Chicago, Chicago, IL 60637}
\date{\today}

\begin{abstract}
Sorting the integers 1 through $N$ into an ordered list is a simple task that can be done rapidly. However, using an algorithm based on the thermally-activated pairwise exchanges of neighboring list elements, we find sorting can display many features of a glass, even for lists as small as $N = 5$. This includes memory and rejuvenation effects during aging --- two hallmarks of glassy dynamics that have been difficult to reproduce in standard glass simulations. 
\end{abstract}
\maketitle

The slow dynamics of some low-temperature, out-of-equilibrium  systems can be highly complex. Without the time nor the ability to sample all of phase space as the dynamics slows down, the relaxation develops a non-exponential form. The system is no longer stationary, but will age, and it will often retain a memory of its previous thermal history. Structural glasses, formed by supercooling liquids, and spin glasses, formed by freezing dilute random spin systems, are quintessential examples with such behavior. Models with simplified dynamics, such as kinetically constrained models \cite{RitortSollich2003}, or those based on parking \cite{BenNaim&c1998, KolanNowakTkachenko1999, TarjusViot2004}, or relaxation on hypercubes \cite{Campbell}, etc., have been proposed to illustrate one or another specific aspect of glassy relaxation. However, even such simplified systems have strained the capacity of computers to reproduce many of the effects readily observed in the laboratory. Here we present a simple algorithmic model, based on sorting a small list of numbers, which displays many of the dynamical features associated with glasses, including those that have resisted replication by large-scale computer simulations. 

Consider the integers $1\dots N$, randomly arranged into a list $S=\{s_1, s_2, \dots, s_N\}$, to be sorted in ascending or descending order. We model sorting as an activated process governed by nearest-neighbor interactions, and define the following Hamiltonian, whose ground state is a sorted list:
\begin{equation}
\begin{split}
H[S] =& \sum_{k=1}^{N-1} \left[ \left(s_{k+1}-s_{k}\right)^{2} - 1\right] - g \sum_{k=1}^{N} \left(k s_{k} - \bar{s}^{2} \right) \\ \equiv & \ H_0 - gM.
\end{split}
\end{equation}
Here $H_{0}$ minimizes the difference between adjacent list elements, and is symmetric between $S$ and its reverse. This symmetry is broken by the second term $gM$. Applying $g>0$ ($g<0$) selects the list sorted in ascending (descending) order as the unique ground state. Subtracting the constant $\bar{s}^{2} = (N+1)/2$ is so that a list and its reverse have values of $M$ that are equal in magnitude but opposite in sign. For a sorted list, $H_{0}=0$ and $M$ has maximum magnitude. The thermal-sorting algorithm (``thermosort'') is as follows: at each time step $t$, we randomly select two adjacent list elements and attempt to swap their places: $S = \{\dots, s_{k}, s_{k+1}, \dots\} \rightarrow S' = \{ \dots, s_{k+1}, s_{k}, \dots \}$.
The probability that this swap will be accepted is 
\begin{equation}
p(S\rightarrow S') = \min\left\{ 1, e^{-(H[S']-H[S])/T}  \right\},
\nonumber
\end{equation}
where $T$ is an effective temperature. Thus $S$ explores a space of $N!$ configurations via the thermally-activated swapping of adjacent list elements. 

\begin{figure}[tb]
\includegraphics[width= 0.45\textwidth]{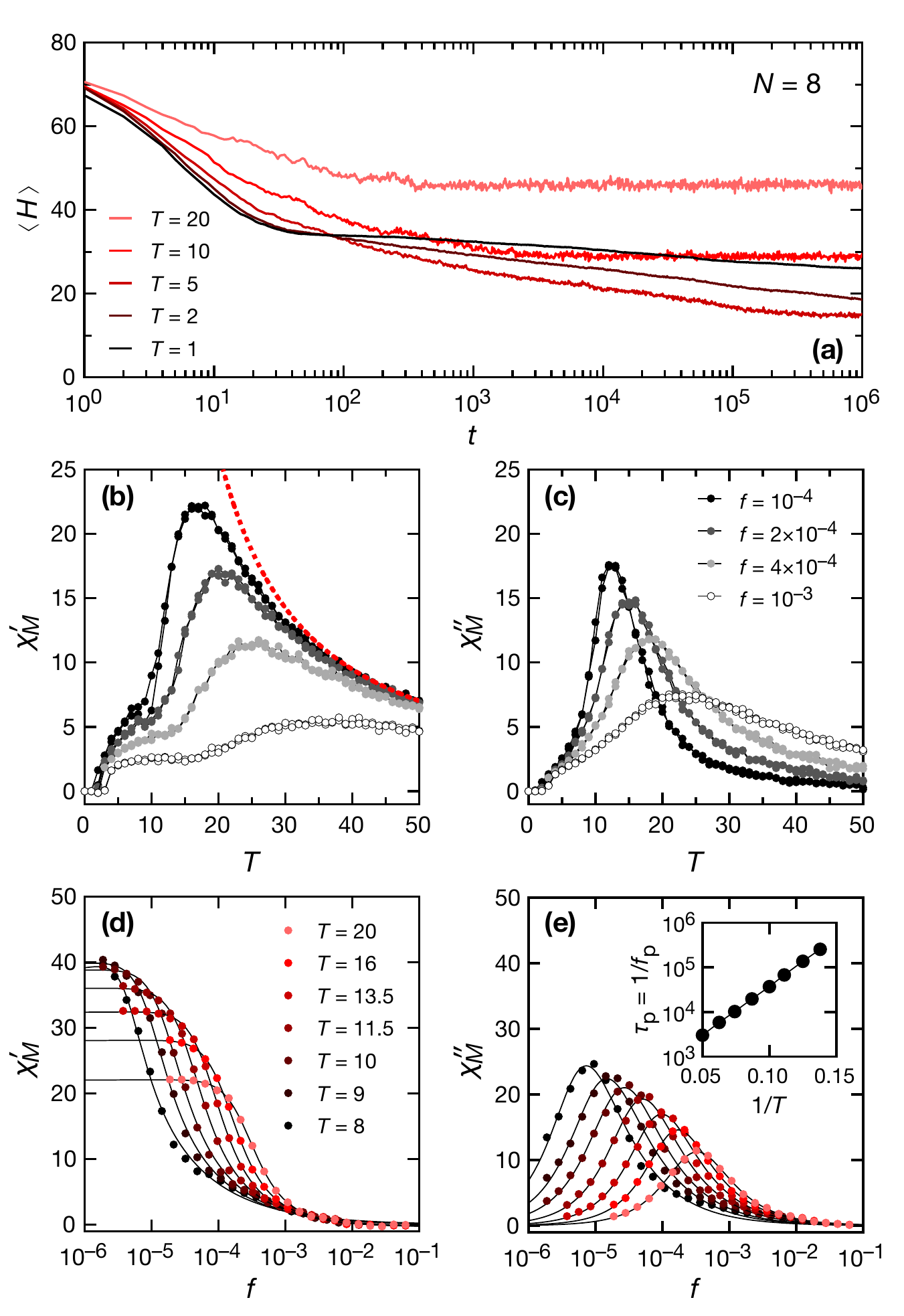}
\caption{(color online) (a) Relaxation of the average energy $\langle H\rangle$, starting from random configurations, at various $T$. (b, c) The complex susceptibility $\tilde{\chi}_{M}$ versus $T$, measured by ramping from $T = 50$ to $T = 0$ and back in steps of  $\Delta T = 1$, and dwell time/step $\tau_\mathrm{dw} = 10^4$. Dashed line in (b) indicates the equilibrium $\chi_{M}$ in the DC limit. (d, e) The equilibrium spectrum of $\tilde\chi_{M}$ at different $T$. Lines are Havriliak-Negami fits \cite{HavriliakNegami1966}. Inset of (e) plots the inverse peak frequency $1/f_\mathrm{p} =\tau_\mathrm{p}$ of $\chi_{M}''(f)$ versus $1/T$. Data shown are ensemble-averages over $10^{3}$ independent realizations for $N=8$ and $g=0$.}
\label{glassy}
\end{figure}  

Fig.~\ref{glassy} illustrates several aspects of glassy dynamics shown by this algorithm for small, $N=8$, lists. Fig. \ref{glassy}(a) plots the relaxation of the ensemble-averaged energy $\langle H\rangle$, starting from random initial states with $g=0$. As $T$ decreases, thermosort takes ever longer to reach thermal equilibrium. For sufficiently low $T$, $\langle H\rangle$ relaxes logarithmically over long stretches of time, indicative of a broad spectrum of relaxation times. The system remains far from thermal equilibrium even after millions of time steps. This is reminiscent of the slow, low-$T$, relaxation (or aging) of structural and spin glasses, as well as the logarithmic compaction of a gently-tapped granular packing \cite{NordbladSvedlindhLundgrenSandlund1986, LehenyNagel1998, Knight&c1995,Zou2010}. It is intriguing to find similar relaxation in lists of only 8 numbers. 

We define a susceptibility $\chi_{M} = \delta M/\delta g$, which describes the linear response of $M$ to a small applied field $g$. Figs.~\ref{glassy}(b, c) plot the $T$-dependence of the complex susceptibility $\tilde{\chi}_{M} = (\chi_{M}', \chi_{M}'')$, measured for $N=8$ by applying a small oscillatory field $\delta g$ at frequency $f$. Both $\chi_{M}'(T)$, $\chi_{M}''(T)$ are singly-peaked functions that shift to lower $T$ and are taller and more sharply defined for lower frequency. Figs.~\ref{glassy}(d-e) plot the frequency spectrum of $\tilde\chi_{M}$ at thermal equilibrium for different temperatures. As $T$ is lowered, the peak frequency $f_\mathrm{p}$ of $\chi_{M}''(f)$ shifts to lower $f$, indicative of a growing characteristic relaxation time $\tau_\mathrm{p} \equiv 1/f_\mathrm{p}$. Inset of Fig.~\ref{glassy}(d) shows that $\tau_\mathrm{p}$ appears to follow an Arrhenius law: $\tau_\mathrm{p} \sim e^{1/T}$. Again, the phenomenology here is very similar to that found in the magnetic and dielectric spectroscopies of spin glasses and glass-forming liquids \cite{BinderYoung1986, EdigerAngellNagel1996, HavriliakNegami1966, Reich&c1990, QuilliamMengMugfordKycia2008, Menon&c1992}.

\begin{figure}[tb]
\includegraphics[width= 0.32\textwidth]{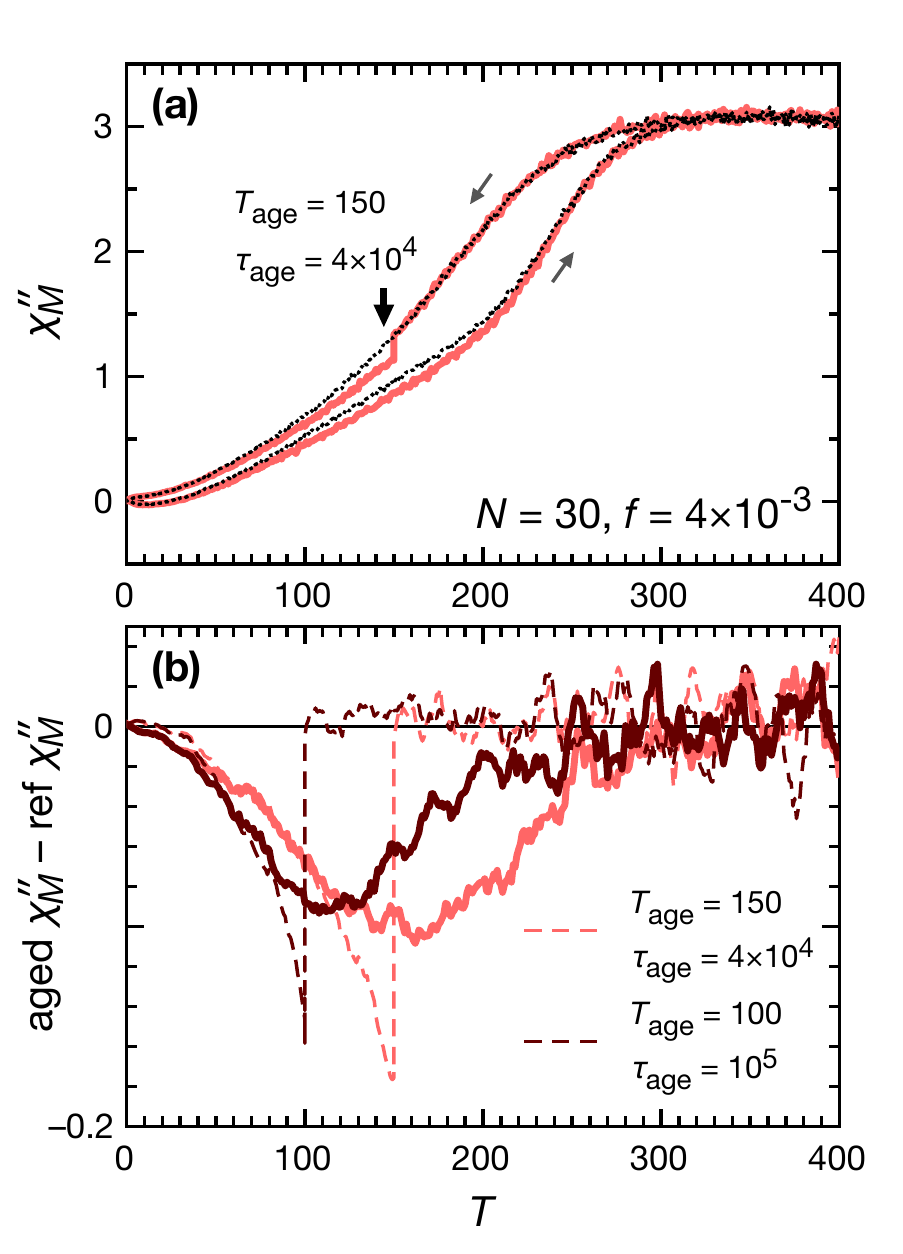}
\caption{(color online) Aging, rejuvenation, and memory in thermosort, measuring $\chi_{M}''(T)$ at $f = 4\times10^{3}$ versus $T$. Here $T$ is ramped in steps of $\Delta T/\tau_\mathrm{dw} = 1/10^{3}$. (a) Dotted curve: reference $\chi_{M}''(T)$ without aging. Solid curve: cooling was interrupted at $T_\mathrm{age} = 150$ for a duration $\tau_\mathrm{age} = 4\times 10^4$, allowing $\chi_{M}''(T)$ to age, forming a dip. Warming back up, the aged $\chi_{M}''(T)$ systematically dips below the reference in the vicinity of $T_\mathrm{age}$. (b) The difference between the aged and reference $\chi_{M}''(T)$, obtained for two different values of $T_\mathrm{age}$ (dotted line: cooling; solid line: warming). Data shown are ensemble-averaged over $10^{5}$ realizations for $N = 30$ and $g=0$.}
\label{memory}
\end{figure}  

\begin{figure}[tb]
\includegraphics[width= 0.48\textwidth]{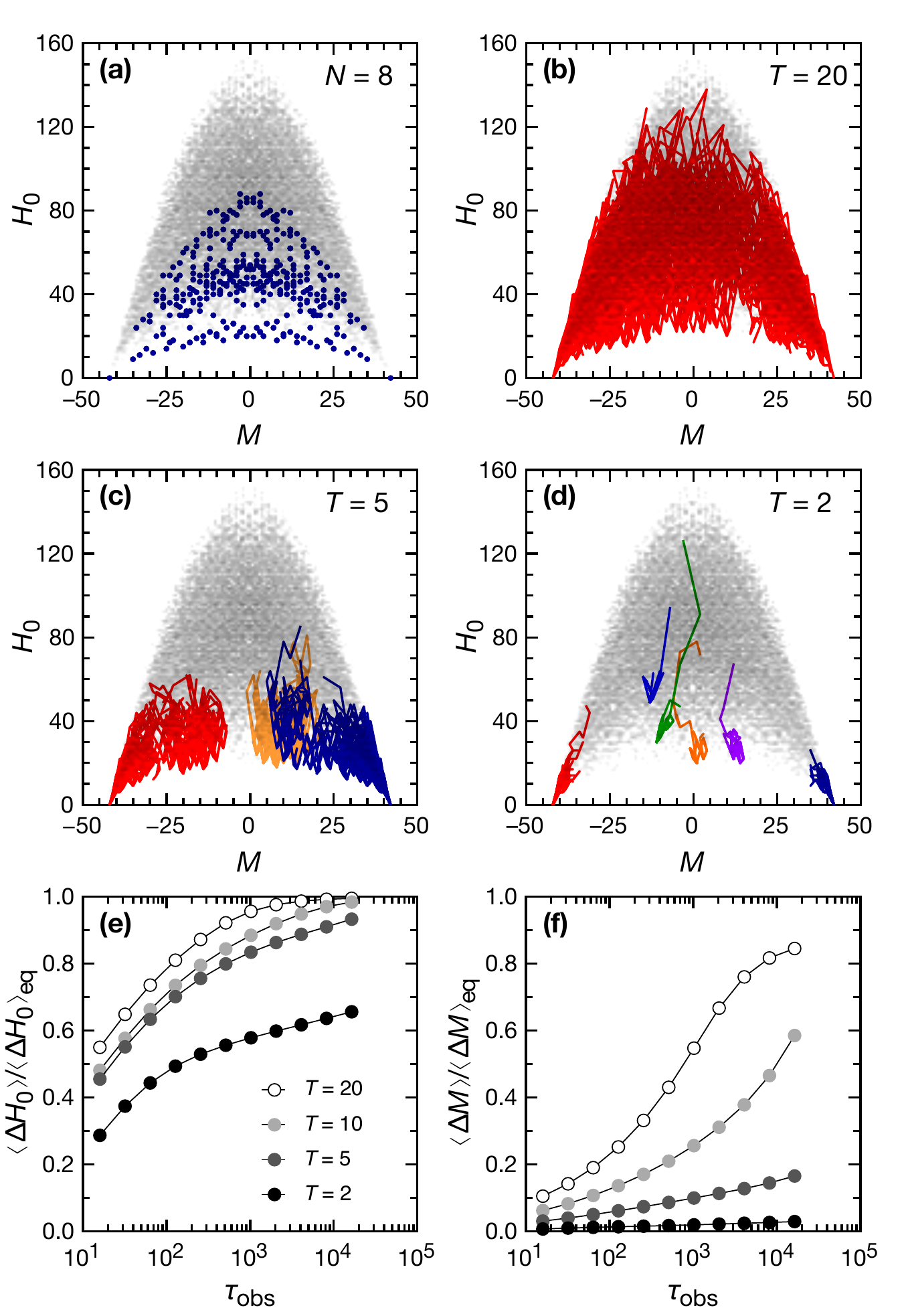}
\caption{(color online) (a) The $N=8$ configuration space projected onto the $(M, H_{0})$ plane for $g=0$; filled circles mark locally-stable states. (b-d) Thermosort trajectories in the projected view, sampled at various $T$. Each color is an independent trajectory, $10^{4}$ in duration, starting from a random configuration. (e, f) Quantifying trajectory localization: the localization extent in $H_{0}$ and and $M$ versus the observational timescale $\tau_\mathrm{obs}$. Here, $\langle\Delta H_{0}(\tau_\mathrm{obs})\rangle$ and $\langle\Delta M(\tau_\mathrm{obs})\rangle$ are ensemble-averaged over 100 independent trajectories, $10^{5}$ in duration; they are then normalized by their equilibrium $\tau_\mathrm{obs}\rightarrow\infty$ values.}
\label{phase_space}
\end{figure}  

At low $T$, the relaxation time of a glassy system is longer than the experimental timescale, and the glass is out of equilibrium. The properties of a glass are thus not stationary but depend on the experimental timescale (aging), and on the thermal history (memory). A particularly striking manifestation of these features is found in the ``rejuvenation and memory'' phenomenon, first observed experimentally in the magnetic susceptibilities of spin glasses, and subsequently in the dielectric susceptibility of organic glass-formers \cite{Jonason&c1998, Dupuis&c2001, BellonCilibertoLaroche2000, YardimciLeheny2003}. If, on cooling a glass from a high-$T$ equilibrium state, the cooling is halted at $T=T_\mathrm{age}$ and kept there for a period $\tau_\mathrm{age}$, then as the glass ages, the (magnetic or dielectric) susceptibility, $\chi''$ slowly decreases in magnitude. When cooling is resumed, after some time, the system appears to forget the aging experience and $\chi''(T)$ reverts to the reference behavior it would have exhibited had the cooling been uninterrupted: the system is said to be ``rejuvenated''. If the glass is subsequently reheated, $\chi''(T)$ initially follows the reference curve. However, as $T$ approaches $T_\mathrm{age}$,  $\chi''(T)$ will mirror the aging-induced dip --- it remembers its cooling history. While seen in many experiments and subjected to several theoretical models, this phenomenon has proven difficult, if not impossible, to observe unambiguously even in sophisticated spin-glass simulations \cite{Jonsson&c2004, Vincent2007, ThomasWhiteMiddleton2008, PiccoRicciTersenghiRitort2001, BerthierBouchaud2002, BerthierYoung2005, JimenezMayorGaviro2005, KrzakalaRicciTersenghi2006}. 

Fig.~\ref{memory} shows thermosort can clearly reproduce memory and rejuvenation, using only $N=30$. Fig.~\ref{memory}(a) shows $\chi_{M}''(T)$  (averaged over many realizations) both in the reference behavior, when cooling is continuous, and when cooling was interrupted at $T_\mathrm{age}$ and the system is allowed to age. In order to see the results more clearly, Fig.~\ref{memory}(b) plots the difference between the aged and reference $\chi_{M}''(T)$ curves for two different values of $T_\mathrm{age}$. The aging dip stands out clearly, and at low temperatures there is rejuvenation as the aging curve reverts to the reference one. On reheating, the memory dip is recovered, and it is clear that the memory dip tracks $T_\mathrm{age}$. Other less complex glassy effects, such as memory/annealing after step-wise shifts in $T$, the Kovacs effect, thermoremanent magnetization, and irreversible/reversible dynamics (as found in granular compaction \cite{Nowak&c1997}), etc., can be easily observed using $N$ as small as 5 \cite{ZouNagelunpub}. 

While Figs.~\ref{glassy} and \ref{memory} describe the average behavior of a large ensemble of independent thermosort realizations, we can also visualize individual ``trajectories''. We do so by projecting the space of $N!$ list configurations onto the $(M, H_{0})$ plane [Fig.~\ref{phase_space}(a)]. While this projection is not unique in that multiple configurations may have the same coordinate, the results are quite illuminating. At high $T$, a trajectory quickly covers the accessible configuration space, as shown in Figure~\ref{phase_space}(b). But as $T$ is reduced, the trajectories become confined over long periods of time to one of a few large basins with only occasional transitions between them, as shown in Fig.~\ref{phase_space}(c). As $T$ is reduced further, the trajectories become localized in smaller, more numerous, basins as shown in Fig.~\ref{phase_space}(d). Thus as $T$ is reduced, the accessible configuration space breaks up into a succession of ever smaller and ever more numerous basins. 

To quantify the extent of localization, we divide trajectories into blocks of length $\tau_\mathrm{obs}$. Within each block, we calculate the standard deviations of $H_{0}$ and $M$ sampled by the trajectory. These standard deviations, $\Delta H_{0}$ and $\Delta M$, are then averaged over all time blocks, and over an ensemble of randomly initiated trajectories. The results $\langle\Delta H_{0}(\tau_\mathrm{obs})\rangle$, $\langle\Delta M(\tau_\mathrm{obs}) \rangle $ describe the localization extent in $H_{0}$ and in $M$ as functions of the observation time $\tau_\mathrm{obs}$. While both quantities grow with $\tau_\mathrm{obs}$, Figs.~\ref{phase_space}(e, f) show that at low temperatures, $\langle\Delta H_{0}\rangle$ grows much faster than $\langle\Delta M\rangle$. This suggests that $M$ is the principal ``coordinate'' along which ergodicity is broken at low $T$. 

\begin{figure*}[tb]
\includegraphics[width= 0.96\textwidth]{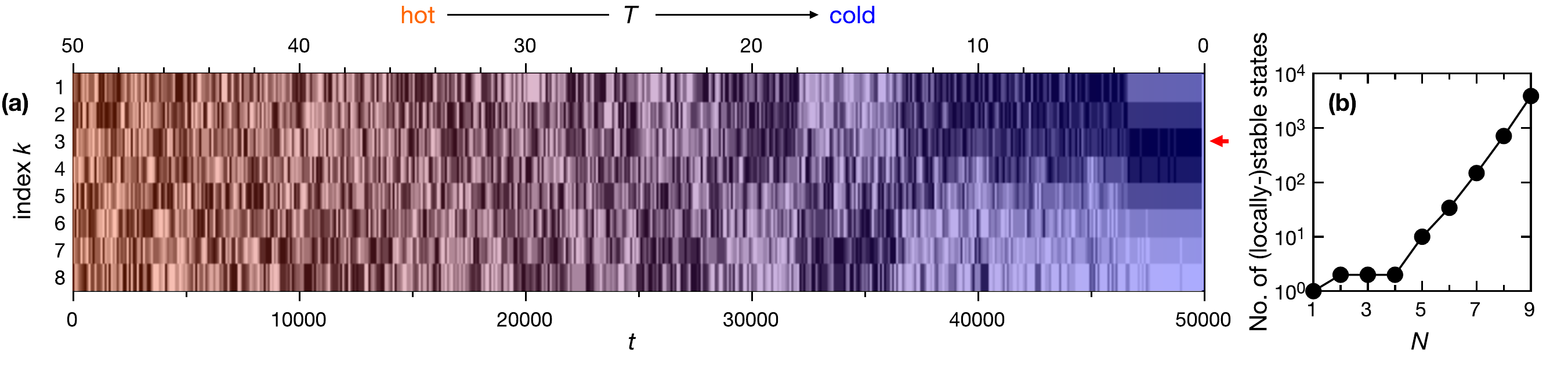}
\caption{(color online) (a) The trajectory of a randomly initiated $N=8$ list ($\textrm{dark}\rightarrow\textrm{light} = 1\rightarrow 8$), slowly cooled from $T=50$ to $T=0$ with $g=0$. The end state at $T=0$ is a locally-stable configuration, consisting of two sorted domains concatenated at a ``domain wall'' ($k=3$, arrow). (b) The number of locally-stable states versus $N$, obtained by enumeration up to $N=9$.}
\label{trajectory}
\end{figure*}  

To understand why trajectories localize in basins at low $T$, Fig.~\ref{trajectory}(a) visualizes the evolving configurations of a randomly initialized $N = 8$ list as it is cooled from $T = 50$ to $T = 0$. The final $T = 0$ state, $S = \{5, 3, 1, 2, 4, 6, 7, 8 \}$ is not fully-sorted, but consists of two sorted ``domains'' $\{5, 3, 1\}$ and $\{1, 2, 4, 6, 7, 8\}$, joined by a ``domain wall'' at $s_{3} = 1$. This configuration is locally-stable: swapping any adjacent pair leads to higher energy. All locally-stable lists have this partially-sorted domain structure; for large $N$, locally-stable configurations can have many domains, so long as each domain contains at least three elements. From a complete enumeration of all lists up to $N = 9$, we find that above $N=5$ the number of locally-stable configurations appears to grow exponentially with $N$ as shown in Fig.~\ref{trajectory}(b), and that the average energy barrier to escape from locally-stable configurations grows at least as rapidly as $N$. These barriers stabilize the domains over long periods of time, allowing even small lists to exhibit glassy signatures. The smallest list in which domains are stable, $N = 5$, is also the smallest $N$ to exhibit glassy dynamics. 

While inside a domain the list is ordered, the numbers are not necessarily in consecutive order. Therefore in order to anneal away a domain, rearrangements must occur throughout and not just at its boundary. Thus even a partially-sorted list with large domains and low energy is configurationally far away from the two fully sorted ground states. This conflict between local and global ordering may be generic in glassy relaxation. For example, it is reminiscent of the proposal that glass formation in simple liquids arises from the incompatibility of local icosahedral order with long-range periodic packing\cite{Nelson2002}. It also suggests that the glassy dynamics of thermosort may be eased by incorporating ``non-local'' swaps into its kinetics to destabilize locally-sorted domains. We define an interaction range $r$, such that $s_{k}$ and $s_{k'}$ can be swapped if $\vert k - k' \vert \leq r$. Setting $r = N - 1$ removes all the locally-stable configurations except the ground states; this allows the system to quickly reach thermal equilibrium even at $T = 0$. Varying $r$ does not change the equilibrium properties of thermosort; but it does speed up the kinetics by increasing the connectivity of the configuration space. This is reminiscent of kinetically constrained glassy models, whose Hamiltonians are often trivial, but can exhibit glassy dynamics when their kinetics are sufficiently constrained \cite{RitortSollich2003}. 

As a sorting algorithm, thermosort is exceedingly poor; but it turns out to be an excellent glass-former. It exhibits a broad array of glassy phenomena, including rejuvenation and memory, which have proven difficult to reproduce in standard simulations of supercooled liquids or spin glasses. Since thermosort is such a simple model, yet capable of exhibiting glassy dynamics with as few as $N=5$ list elements, we suspect it probably contains little more than the minimal makings of a glassy system. 

We thank Susan Coppersmith, Florent Krzakala, Erika Nesse, Yair Shokef and Lenka Zdeborov\'a for stimulating discussions and helpful suggestions. This work was supported by NSF grant DMR-0652269.


\begin{thebibliography}{10}
\bibitem{RitortSollich2003}F. Ritort and P. Sollich, {Adv. Phys.} \textbf{52}, 219 (2003).
\bibitem{BenNaim&c1998}E. Ben-Naim, J. B. Knight, E. R. Nowak, H. M. Jaeger, and S. R. Nagel, {Physica D} \textbf{123}, 380 (1998).
\bibitem{KolanNowakTkachenko1999}A. J. Kolan, E. R. Nowak and A. V. Tkachenko, {Phys. Rev. E} \textbf{59}, 3094 (1999).
\bibitem{TarjusViot2004}G. Tarjus and P. Viot, {Phys. Rev. E} \textbf{69}, 011307 (2004).
\bibitem{Campbell}I. A. Campbell, J. M. Flesselles, R. Jullien, and R. Botet, {Phys. Rev. B} \textbf{37}, 3825 (1988).
\bibitem{NordbladSvedlindhLundgrenSandlund1986}P. Nordblad, P. Svedlindh, L. Lundgren, and L. Sandlund, {Phys. Rev. B} \textbf{33}, 645 (1986). 
\bibitem{LehenyNagel1998}R. L. Leheny and S. R. Nagel, {Phys. Rev. B} \textbf{57}, 5154 (1998).
\bibitem{Knight&c1995}J. B. Knight, C. G. Fandrich, C. N. Lau, H. M. Jaeger, and S. R. Nagel, {Phys. Rev. E} \textbf{51}, 3957 (1995).
\bibitem{Zou2010} L.-N. Zou, {Phys. Rev. E} \textbf{81}, 031302, (2010).
\bibitem{BinderYoung1986}K. Binder and A. P. Young, {Rev. Mod. Phys.} \textbf{58}, 801 (1986).
\bibitem{EdigerAngellNagel1996}M. D. Ediger, C. A. Angell, and S. R. Nagel, {J. Phys. Chem.} \textbf{100}, 13200 (1996).
\bibitem{HavriliakNegami1966}S. Havriliak and S. Negami, {J. Polymer Sci. Part C} \textbf{14}, 99 (1966).
\bibitem{Reich&c1990}D. H. Reich, B. Ellman, J. Yang, T. F. Rosenbaum, G. Aeppli, and D. P. Belanger, {Phys. Rev. B} \textbf{42}, 4631 (1990).
\bibitem{QuilliamMengMugfordKycia2008}J. A. Quilliam, S. Meng, C. G. A. Mugford, and J. B. Kycia, {Phys. Rev. Lett.} \textbf{101}, 187204 (2008).
\bibitem{Menon&c1992}N. Menon, K. P. O'Brien, P. K. Dixon, L. Wu, S. R. Nagel, B. D. Williams, and J. P. Carini, {J. Non-Cryst. Solids} \textbf{141}, 61 (1992).
\bibitem{Jonason&c1998}K. Jonason, E. Vincent, J. Hammann, J.-P. Bouchaud, and P. Nordblad, {Phys. Rev. Lett.} \textbf{81}, 3243 (1998).
\bibitem{Dupuis&c2001}V. Dupuis, E. Vincent, J.-P. Bouchaud, J. Hammann, A. Ito, and H. A. Katori, {Phys. Rev. B} \textbf{64}, 174204 (2001).
\bibitem{BellonCilibertoLaroche2000}L. Bellon, S. Ciliberto, and C. Laroche, {Europhys. Lett.} \textbf{51}, 551 (2000).
\bibitem{YardimciLeheny2003}H. Yardimci and R. L. Leheny, {Europhys. Lett.} \textbf{62}, 203 (2003).
\bibitem{ThomasWhiteMiddleton2008}C. K. Thomas, O. L. White, A. A. Middleton, {Phys. Rev. B} \textbf{77},  092415 (2008).
\bibitem{Vincent2007}E. Vincent, \textit{Ageing and the Glass Transition}, edited by M. Henkel, M. Pleimling, and R. Sanctuary (Springer, Berlin, 2007).
\bibitem{Jonsson&c2004}P. E. J\"onsson, R. Mathieu, P. Nordblad, H. Yoshino, H. A. Katori, and A. Ito, {Phys. Rev. B} \textbf{70}, 174402 (2004).
\bibitem{PiccoRicciTersenghiRitort2001}M. Picco, F. Ricci-Tersenghi, and F. Ritort, {Phys. Rev. B} \textbf{63}, 174412 (2001).
\bibitem{BerthierBouchaud2002}L. Berthier and J.-P. Bouchaud, {Phys. Rev. B} \textbf{66}, 054404 (2002).
\bibitem{BerthierYoung2005}L. Berthier and A. P. Young, {Phys. Rev. B} \textbf{71}, 214429 (2005)
\bibitem{JimenezMayorGaviro2005}S. Jim\'enez, V. Mart\'in-Mayor, and S. P\'erez-Gaviro, {Phys. Rev. B} \textbf{72}, 054417 (2005).
\bibitem{KrzakalaRicciTersenghi2006}F. Krzakala and F. Ricci-Tersenghi, J. Phys: Conf. Ser. \textbf{40}, 42 (2006).
\bibitem{Nowak&c1997}E. R. Nowak, J. Knight, M. Povinelli, H. M. Jaeger, and S. R. Nagel, {Powder Technol.} \textbf{94}, 79 (1997).
\bibitem{ZouNagelunpub}L.-N. Zou and S. R. Nagel, in preparation.
\bibitem{Nelson2002}D. R. Nelson, \textit{Defects and Geometry in Condensed Matter Physics} (Cambridge University Press, New York, 2003).

\end{thebibliography}
\end{document}